\documentclass[11pt]{article}
\usepackage{moriond,epsfig}

\def\Journal#1#2#3#4{{#1} {\bf #2}, #3 (#4)}

\def\NIMA{{\em Nucl. Instrum. Methods} A}

\def\PLB{{\em Phys. Lett.}  B}
\def\PRL{\em Phys. Rev. Lett.}
\def\PRD{{\em Phys. Rev.} D}

\def\PR{{\em Phys. Rev.}}

\def\be{\begin{equation}}
\def\ee{\end{equation}}
\def\bea{\begin{eqnarray}}
\def\eea{\end{eqnarray}}

\begin{document}
\vspace*{4cm}
\title{PRECISION MEASUREMENTS IN NEUTRON DECAY}


\author{M. SCHUMANN\\ for the PERKEO II collaboration}
\address{Physikalisches Institut, Universit\"at Heidelberg, Philosophenweg 12, 69120 Heidelberg, Germany}

\maketitle\abstracts{We present new precision measurements of angular correlation coefficients
in polarized neutron decay. We have obtained values for the electron asymmetry coefficient $A$, the 
neutrino asymmetry coefficient $B$, and for the proton asymmetry coefficient $C$. In combination with 
other results, the new measurements are used to derive limits on ``Physics beyond the Standard Model".}

\section{Introduction}

Free neutrons decay with a mean lifetime $\tau_n$ of about 15 minutes in electron,
proton, and electron anti-neutrino: $n \to e^- \ p \ \overline{\nu}_e$.
The maximal kinetic energy of the electron is $E_\textnormal{\footnotesize max}(e)$ = 782 keV. For the proton, 
$E_\textnormal{\footnotesize max}(p)$ = 780 eV is three orders of magnitude smaller. 
Neutron decay -- involving all
particles of the first generation -- constitutes an ideally suited laboratory to do
particle physics at very low energies as it provides important information
on the structure of the weak interaction an the underlying symmetries. It
is a quite simple system where theoretical corrections are small 
and no nuclear structure effects have to be considered.

Our observables are angular correlations between neutron spin and the momentum of
the three decay products. The decay probability d$\omega$ of polarized neutrons
can be expressed in terms of two of these ``asymmetries'', the
electron asymmetry $A$ and the neutrino asymmetry $B$ \cite{Jac57}:
\be \label{jackson} 
\textnormal{d} \omega \propto |V_{ud}|^2 \left( 1 + a \ \frac{\boldmath{p}_e \boldmath{p}_\nu}{E E_\nu} + \langle \boldmath{s}_n
\rangle \left[ A \ \frac{\boldmath{p}_e}{E} + B \ \frac{\boldmath{p}_\nu}{E_\nu}\right]\right) \textnormal{.}
\ee
$E$, $E_\nu$, $\boldmath{p}_e$, and $\boldmath{p}_\nu$ are energy and momentum of electron
and neutrino respectively, $\langle \boldmath{s}_n \rangle$ denotes the neutron spin, and
$|V_{ud}|$ is the first entry of the quark mixing matrix.
$a$ is the correlation between electron and neutrino momentum. The proton asymmetry $C$
does not enter this expression, but it is kinematically coupled to $A$ and $B$ via \cite{Glu96}
$C = x_C (A+B)$, where $x_C=0.27484$ is a kinematical factor. In the standard $V-A$
formulation of weak interactions, all correlations are functions of one single 
parameter $\lambda = g_A/g_V$, the ratio of axial-vector and vector coupling constant
(assuming $\lambda$ to be real and neglecting weak magnetism and other recoil effects):
\be\label{eq_correlations}
a=\frac{1-\lambda^2}{1+3\lambda^2}\textnormal{,} \quad A=-2\ \frac{\lambda^2 +
\lambda}{1+3\lambda^2}\textnormal{,} \quad B=2 \ \frac{\lambda^2-\lambda}{1+3\lambda^2}
\textnormal{,} \quad  C = x_C \ \frac{4 \ \lambda}{1+3 \lambda^2} \textnormal{.}
\ee
The precise determination of $\lambda$ is important as the calculation of
many processes in cosmology (e.g. primordial element formation), astronomy (e.g. solar cycle, 
neutron star formation), and particle physics (e.g. neutrino detectors, neutrino scattering)
depends on this parameter. The observable most sensitive to $\lambda$ is the electron asymmetry $A$,
however, it can be also extracted from measurements of $a$ and $C$.

Within the Standard Model, neutron decay can be described with the two parameters $\lambda$ and
$V_{ud}$ only. But since much more observables are accessible (various correlation coefficients and the
lifetime $\tau_n$) the problem is overdetermined, and neutron decay can be used to test the
Standard Model and to search for new physics. In section \ref{sec_New}, we will present neutron
limits for additional right-handed ($V+A$) currents and anomalous couplings (scalar and tensor)
in the interaction.

\section{Experiment and Results}

\begin{figure}[b!]
\begin{center}
\epsfig{figure=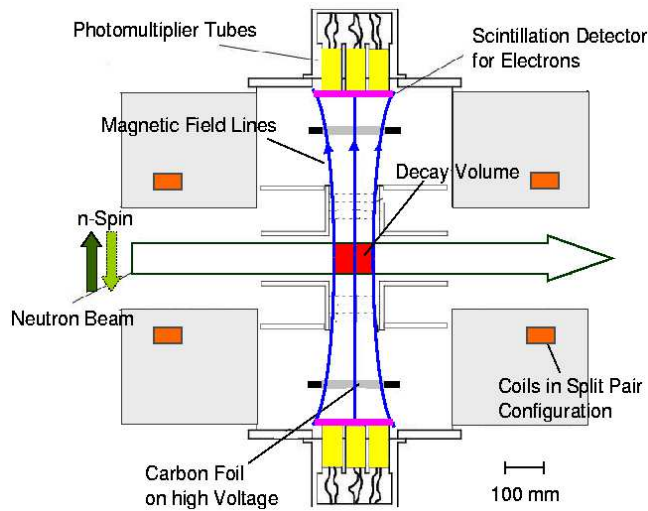,width=8.5cm}
\end{center}
\caption{\label{fig_Perkeo} Electron spectrometer PERKEO II: Transversally polarized
neutrons transit the instrument, their spins are aligned with the magnetic field, which separates the
full solid angle into two hemispheres covered by a detector each.}
\end{figure}

The measurements were performed using the electron spectrometer PERKEO II. It features a
pair of superconducting coils in a split pair configuration that generate a
slightly decreasing magnetic field ($B_\textnormal{\footnotesize max} \approx$ 1 T)
perpendicular to the spin polarized neutron beam crossing
the spectrometer (cf. fig. \ref{fig_Perkeo}). The instrument was installed at the
cold neutron beam position PF1B at the Institut Laue-Langevin (ILL), Grenoble.

The magnetic field fulfills several functions: It separates the full solid angle into
two hemispheres since the spins align with the field lines: The momentum projection of
the decay particle onto the neutron spin determines whether the particle is emitted in
neutron spin direction or against it.
The field guides the charged decay products onto the
two detectors, installed at both sides next to the beam, providing full 2 $\times$ 2
$\pi$ detection. The detectors consist of plastic scintillators with photomultiplier
readout. In order to measure the very low energetic protons with the same detector 
(necessary for neutrino and proton asymmetry), we developed a special setup to convert protons
into electrons \cite{Kre05b,Sch07}: Protons were accelerated onto a very thin carbon foil
on negative high voltage where they had enough ionization power to generate
one or more secondary electrons. These could be detected by the scintillators.

\subsection{Electron Asymmetry $A$ and $\lambda$}

The experimental signature of the electron asymmetry is 
\be\label{eq_A}
A_\textnormal{\footnotesize exp}(E) = \frac{N^+(E)-N^-(E)}{N^+(E)+N^-(E)} =  \frac{1}{2} \frac{v}{c} A \ P \ F\textnormal{,}
\ee
where $N(E)$ is the number of electrons with energy $E$ and the sign denotes whether
the electron was emitted in ($+$) or against ($-$) neutron spin direction. This
expression holds for both detectors separately since a spinflipper was 
used to periodically turn the neutron spin by 180$^\circ$. It is related to the angular 
correlation coefficient $A$ via the second equation in (\ref{eq_A}). 
$v/c$ is the electron velocity in terms of the speed of light, $P$ the neutron
polarization, and $F$ the spinflipper efficiency. With new methods to
polarize the beam and to analyze the beam polarization \cite{Kre05}
we managed to achieve a much improved beam polarization compared to former measurements.

Background generated in the neutron collimation system and at the beamstop at the end
of the installation was heavily suppressed. The remaining background leads to a small
uncertainty of 0.1 \%. For future measurements, the spectrometer PERKEO III \cite{Mae06} 
has been developed: It can be operated with
a chopped neutron beam allowing to acquire data only when no background is generated.

About 160 million events were recorded and the measurement is still limited by statistics.
Details on the experiment can be found in \cite{Mun06}. For the following analysis, we
will use the value $A_\textnormal{\footnotesize PII} = -0.1193(4)$ -- corresponding
to $\lambda_\textnormal{\footnotesize PII} = -1.2749(11)$, the average
of the new preliminary result and former PERKEO II measurements \cite{Abe02}.

\subsection{Neutrino Asymmetry $B$}

In our setup, the neutrino could not be detected directly, thus it had to be reconstructed from a
coincident measurement of electron and proton. The case where both
are emitted into the same hemisphere is most sensitive to the
neutrino asymmetry $B$ \cite{Glu95}. Here, the experimental asymmetry is defined using 
the electron spectra $Q^{ij}$
\be
B_\textnormal{\footnotesize exp}(E) = \frac{Q^{--}(E)-Q^{++}(E)}{Q^{--}(E)+Q^{++}(E)} \textnormal{,}
\ee
where the first sign indicates the emission direction of the electron, the second denotes the proton.

\begin{figure}[t]
\begin{center}
\epsfig{figure=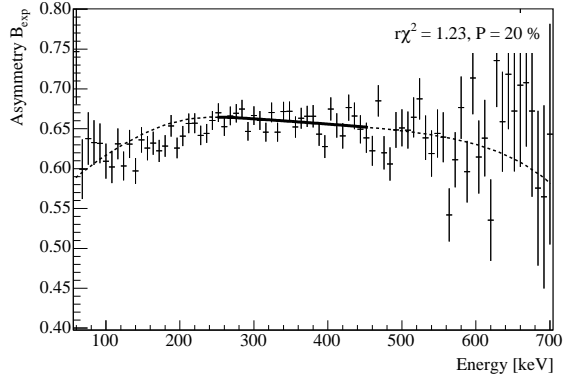,width=7.5cm}
\end{center}
\caption{\label{fig_BFit} Experimental neutrino asymmetry $B_\textnormal{\tiny exp}$. 
The solid line indicates the fit region, the dashed line the fit extension. The result
does not depend on the fit region. Increasing the region to higher energies would increase
the systematic error due to the magnetic mirror effect.}
\end{figure}

A fit to the combined data is shown in figure \ref{fig_BFit}. The result \cite{Sch07}
\be
B_\textnormal{\footnotesize PII} = 0.9802(50)
\ee
is almost independent from detector calibation due to the flat characteristics of the spectrum. 
It has an uncertainty that is comparable to the most precise measurement so far \cite{Ser98}
but has significantly lower corrections. The result is limited by statistics and the error in
the relative position between the magnetic field of the spectrometer and 
the neutron beam maximum. A misalignment would cause some charged particles to be reflected 
at the increasing magnetic field (``magnetic mirror'') leading to signals in
the wrong detector.

The new result agrees with the Standard Model expectation (calculated with eq. (\ref{eq_correlations}) 
and the current PDG value for $\lambda$) and previous measurements \cite{Ser98,Kuz95}. Averaging all
yields the new world mean value with an uncertainty lowered by 25 \%:
\be \label{eq_Bmean}
B_\textnormal{\footnotesize mean} = 0.9807(30) \textnormal{.}
\ee

\subsection{Proton Asymmetry $C$}

The combined electron-proton detector only allows to measure the emission direction
of the proton, a determination of its energy is impossible since it is much smaller than
the electron energy. Therefore we have
to obtain the proton asymmetry $C$ from the coincident measurement
in an integral way. For both detectors, altogether four electron energy 
spectra generated with certain conditions on the emission direction of electrons (first) and
protons (second sign) are available:
\be
Q^{++}(E) \textnormal{,} \quad Q^{-+}(E) \textnormal{,} \quad Q^{--}(E) \textnormal{,} \quad Q^{+-}(E) \textnormal{.}
\ee
Now, the proton asymmetry is defined using the integrals of these spectra
\be
C = \frac{\int (Q^{++}(E)+Q^{-+}(E)) \textnormal{d}E - \int (Q^{--}(E)+Q^{+-}(E)) \textnormal{ d}E}{\int (Q^{++}(E)+Q^{-+}(E)) \textnormal{d}E + \int (Q^{--}(E)+Q^{+-}(E)) \textnormal{d}E} \textnormal{.}
\ee
The integrals were obtained from one-parameter fits to the $Q^{ij}$ spectra in a
fit region that was well above the detection threshold and remaining background
contributions. After extrapolation to the whole energy range, the functions were
integrated. This yields the result \cite{Sch07b}
\be
C_\textnormal{\footnotesize PII} = -0.2377(26) \textnormal{.}
\ee
With an uncertainty of 1.1 \%, this constitutes the first precision measurement of
this observable. The error is dominated by energy calibration and extrapolation, however,
at the moment the proton asymmetry is known more precisely than the electron-neutrino correlation $a$.

\section{Neutron Decay Limits on New Physics}\label{sec_New}

We use the new results in combination with other measurements to set limits
on possible contributions of new physics to neutron decay. Another important input
parameter is the neutron lifetime $\tau_n$. At the moment, however, the
experimental situation concerning $\tau_n$ is quite unclear, since there is one
new measurement \cite{Ser05} deviating by 6.5 $\sigma$ from the average of previous 
results. In order to account for this discrepancy we enlarge all quoted errors with a
scaling factor 2.5 to obtain statistical agreement ($\chi^2/\textnormal{NDF}=1$) and get the average
$\tau_\textnormal{\footnotesize mean} = 882.0(14)$ s. 

\subsection{Right-Handed Currents}

\begin{figure}[b!]
\begin{minipage}{7.6cm}
\begin{center}
\epsfig{figure=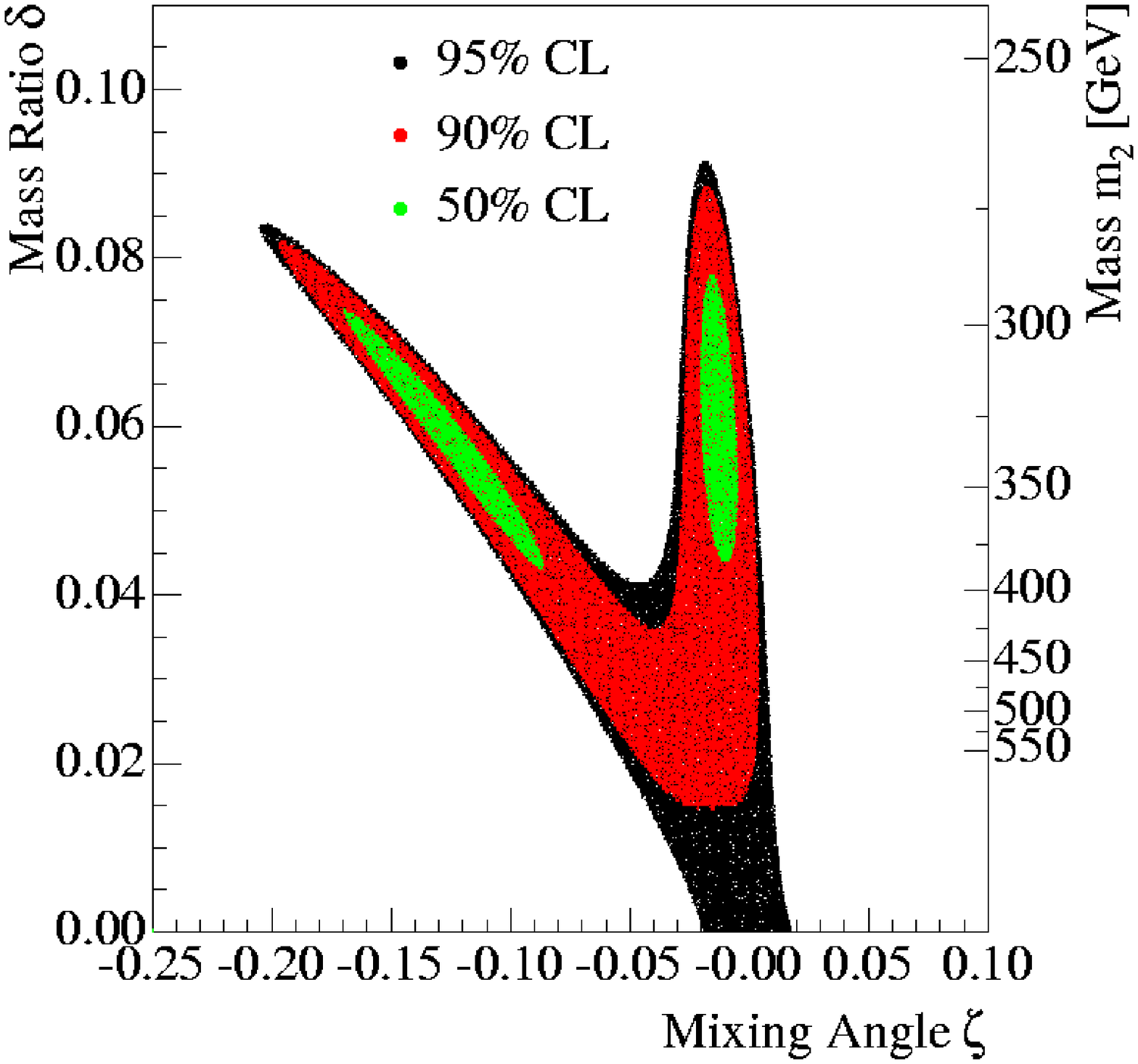,width=6.5cm}
\end{center}
\caption{Exclusion plot on possible admixtures of right-handed ($V+A$) couplings
in the weak interaction derived from neutron decay observables. 
The Standard Model, $\zeta=\delta=0$, is included with 95 \% CL. \label{fig_RHC}}
\end{minipage}
\hfill
\begin{minipage}{7.6cm}
\begin{center}
\epsfig{figure=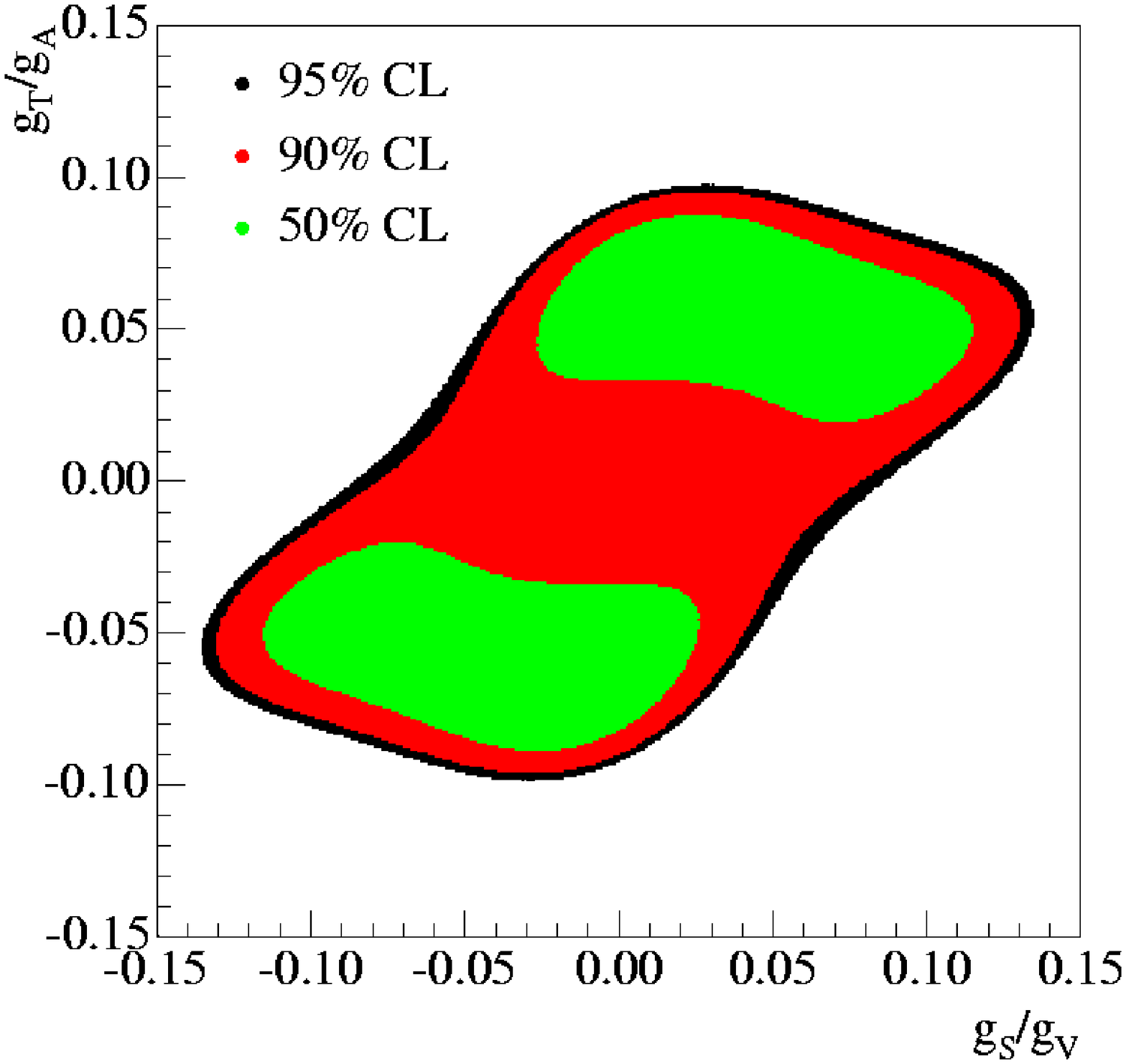,width=6.2cm}
\end{center}
\caption{Neutron decay limits on anomalous (scalar $g_S$ and tensor $g_T$) couplings in 
the weak Lagrangian. The Standard Model $g_S=g_T=0$ is included in the 90 \% CL contour.  \label{fig_CsCt}}
\end{minipage}
\end{figure}

Parity is maximally violated in the Standard Model, i.e. the weak interaction couples
only to left-handed particles. However, the model only describes but 
gives no intrinsic motivation for parity violation. 
According to simple extensions of the Standard Model, the {\it Left-Right Symmetric 
Models} \cite{Pat74,Moh75}, parity violation stems from a spontaneous symmetry breaking at 
mass scale $m_2$. Below this energy, the interaction is mediated by the usual $W_L$-bosons,
however, there should be additional heavy bosons $W_R$,
remnants of a right-handed SU(2)$_R$ group. The {\it Manifest Left-Right Symmetric Model} \cite{Beg77}
assumes that left- and right-handed quark mixing matrices are equal. Here, the weak 
eigenstates $W_{L,R}$ are linear combinations of the mass eigenstates $W_{1,2}$,
\be\label{eq_mixing}
W_L = \cos \zeta \ W_1 - \sin \zeta \ W_2 \qquad \textnormal{and} \qquad W_R = e^{i\phi} \sin \zeta \ W_1 + e^{i\phi} \cos \zeta \ W_2 \textnormal{,}
\ee
with mixing angle $\zeta$ and a $C\!P$ violating phase $\phi$ (we will neglect $\phi$ in 
the following since it has no observable effect). Additional parameters of the theory are
the mass ratio $\delta=m_1^2/m_2^2$ and the coupling constant ratio $\lambda'=g'_A/g'_V$.

If one extends equations (\ref{eq_correlations}) and the expression for the lifetime $\tau_n$ 
to account for possible right-handed admixtures, one can generate exclusion plots and
derive limits. The neutron decay result for the input parameters $A_\textnormal{\footnotesize PII}$, 
$B_\textnormal{\footnotesize mean}$, and $\tau_\textnormal{\footnotesize mean}$ is given 
in figure \ref{fig_RHC} showing a projection along the $\lambda'$ axis onto the $\zeta-\delta$ plane. 
The 90 \% confidence level (CL) limits are $-0.1968 < \zeta < 0.0040$ and $\delta<0.0885$ which
yields $m_2>270$ GeV. 

There are tighter limits on the mass $m_2$ from direct $W'$ searches at 
Tevatron \cite{Ada07}, and better constraints on $\zeta$ from muon decay
tests \cite{Mus05}, however, in the mass range not excluded by the collider
results the neutron limits for the mixing angle $\zeta$ are more stringent. And when one
considers more general left-right symmetric models, results from $\beta$-decay,
muon decay, and direct searches are complementary \cite{Sev06}.

\subsection{Scalar and Tensor Couplings}

In the framework of the Standard Model, the weak interaction includes only vector ($V$) and
axial-vector ($A$) couplings. However, the most general Lorentz invariant Lagrangian \cite{Com83}
\be
\label{eq_generalH}
{\cal L} = \sum_k \ (\overline{p} \Omega_k n) \ (\overline{e} \Omega_k (g_k + g_k' \gamma^5) \nu_e ) + \textnormal{h.c.}
\ee
allows also scalar ($S$) and tensor ($T$) contributions, where the operator $\Omega_k$
($k=V,A,S,T$) describes the kind of interaction. Again, equations 
(\ref{eq_correlations}) were extended to account for additional couplings $g_S$, $g_T$ in 
the simple {\it Right-Handed Scalar and Tensor Model} \cite{Sev06}. It assumes
$g'_V/g_V=1$, $g'_A/g_A=1$, $g'_S/g_V=-g_S/g_V$, and $g'_T/g_A=-g_T/g_A$ leading to
left-handed $V$, $A$ and right-handed $S$, $T$ couplings.
The result of a $\chi^2$-scan based on the input parameters $A_\textnormal{\footnotesize PII}$,
$B_\textnormal{\footnotesize mean}$, $C_\textnormal{\footnotesize PII}$, $a=-0.103(4)$ \cite{PDG06},
and $\tau_\textnormal{\footnotesize mean}$ is shown in fig. \ref{fig_CsCt}. The
90 \% CL limits, $|g_S/g_V|<0.130$ and $|g_T/g_A|<0.0948$, agree with the Standard Model
case $g_S=g_T=0$.

Scalar contributions seem to be almost excluded by an analysis of superallowed
$\beta$-decays giving the constraint $|g_S/g_V|<0.0013$ when the conserved
vector current hypothesis (CVC) is assumed \cite{Har05}. For tensor contributions, however,
neutron decay provides excellent limits.

\section*{Acknowledgments} This work was funded by the German Federal Ministry
for Research and Education unter contract no. 06HD153I.


\end{document}